\documentclass[9pt,twocolumn,twoside]{osajnl}
\usepackage{epstopdf}
\journal{ol} 

\setboolean{shortarticle}{true} 

\ifthenelse{\boolean{shortarticle}}{\colorlet{color2}{color2b}}{\colorlet{color2}{color2}} 

\title{Measuring the electromagnetic chirality of 2D arrays under normal illumination}

\author[1,2,*]{X. Garcia-Santiago}
\author[2,4]{S. Burger}
\author[1,3]{C. Rockstuhl}
\author[1]{I. Fernandez-Corbaton}

\affil[1]{Institute of Nanotechnology, Karlsruhe Institute of Technology, 76021 Karlsruhe, Germany}
\affil[2]{JCMwave GmbH., 12163 Berlin, Germany}
\affil[3]{Institute of Theoretical Solid State Physics, Karlsruhe Institute of Technology, 76131 Karlsruhe, Germany}
\affil[4]{Zuse Institute Berlin, 14195 Berlin, Germany}

\affil[*]{xavier.garcia-santiago@kit.edu}

\dates{Compiled \today}

\ociscodes{(160.1585) Chiral media, (290.5825) Scattering theory, (000.3860) Mathematical methods in physics, (160.3918) Metamaterials}

\doi{\url{http://dx.doi.org/10.1364/ao.XX.XXXXXX}}

\begin{abstract}
We present an electromagnetic chirality measure for 2D arrays of subwavelength periodicities under normal illumination. The calculation of the measure uses only the complex reflection and transmission coefficients from the array. The measure allows the ordering of arrays according to their electromagnetic chirality, allowing a quantitative comparison of different design strategies. The measure is upper bounded, and the extreme properties of objects with high values of electromagnetic chirality make them useful in both near- and far-field applications. We analyze the consequences that different possible symmetries of the array have on its electromagnetic chirality. We use the measure to study four different arrays. The results indicate the suitability of helices for building arrays of high electromagnetic chirality, and the low effectiveness of a substrate for breaking the transverse mirror symmetry.
\end{abstract}

\setboolean{displaycopyright}{true}

\begin{document}

\maketitle
\thispagestyle{fancy}

\ifthenelse{\boolean{shortarticle}}{\ifthenelse{\boolean{singlecolumn}}{\abscontentformatted}{\abscontent}}{}

An object is chiral when it can not be superposed onto its mirror image using any combination of rotations and translations. Although the determination of whether an object is chiral or not is straightforward, the quantification of its degree of chirality is not. Even though many authors have tried to establish a chirality measure \cite{Petitjean:Chirality_measures}, all attempts have encountered substantial problems \cite{Fowler:Quantification_of_chirality}. While the question of how chiral is an object cannot be unambiguously answered, the question of how {\em electromagnetically chiral} is an object can be unambiguously answered. Recently, a general measure for electromagnetic chirality was proposed in \cite{Ivan:Maximum_chirality}. The measure is based on the quantification of the differences in the interaction of the object with fields of different helicity (polarization handedness). This measure is consistent with the geometrical definition of chirality: A geometrically achiral object is always electromagnetically achiral. The measure has an upper bound, and objects that approach the upper bound have extremal light-matter-interaction properties. For example, a maximally electromagnetically chiral object is transparent to all fields, near and far, of a given helicity. This means that their scattered near- and far-fields are of pure helicity, which is useful in applications like polarization control at the nanoscale, sensing of chiral molecules, and angle independent polarization filters (see \cite[Sec. VI]{Ivan:Maximum_chirality}). The electromagnetic chirality of a given object is hence a quantity of interest, allowing a quantitative comparison of different designs, and whose maximization can be used as a design goal. The calculation of the electromagnetic chirality (em-chirality) of a general object is based on the complete knowledge of the interaction of the object with any electromagnetic field. Namely, the complete scattering matrix of the object is needed in order to calculate its em-chirality.

In this work, we develop the em-chirality measure for 2D arrays of subwavelength periodicities under normal illumination. In this setting, the calculation and/or experimental determination of em-chirality is substantially simplified with respect to the general case. We provide a detailed recipe for computing the em-chirality using only the complex reflection and transmission coefficients that are routinely obtained by numerical simulations and/or measurements of the arrays. Our method is valid for planar arrays of arbitrary 3D objects as long as the periodicities of the arrays are smaller than the wavelength of the illumination. We analyze the consequences that different possible symmetries of the array have on its em-chirality. In particular, we find that if the array has any mirror symmetry its em-chirality is zero. Finally, we compute the frequency dependent em-chirality of four different arrays. They represent different strategies for achieving geometric chirality of the arrays. The results indicate that helices are suitable for achieving large em-chirality values and that the presence of a substrate is not an efficient mechanism for breaking the transverse mirror symmetry of planar arrays. The unambiguous ordering of the arrays established by their em-chirality is, for these particular four cases, independent of the frequency. 

The normal illumination of a 2D array with subwavelength periodicities is a common experimental setting. For example, it is used in the characterization of atomically thin materials \cite{PhysRevLett.101.196405}, and artificial materials \cite{Pshenay-Severin:10}. In particular, the design of artificial 2D arrays of inclusions exhibiting chiro-optical responses is receiving lots of research attention \cite{Hentschele,Schaferling:Enhanced_optical_chirality,Lodahl2017,Decker2007,Kaschke2014}. As for any other object, a 2D array can only be chiral if it lacks all mirror symmetries \cite{Fedotov:Asymmetric_propagation_of_em_waves,Menzel:Jones_classification_metamaterials}. This can be achieved in different ways. For example, by exploiting the presence of a substrate to break the transverse mirror symmetry \cite{kuwata:Giant_optical_activity,Powell:Substrate-induced_bianisotropy}, by using chiral inclusions \cite{Semchenko:Chiral_metamaterial_negative_refr,Saenz:Modeling_of_spirals}, or by appropriately stacking achiral objects to form a chiral unit cell \cite{yin:Active_Chiral_Plasmonics,Yin:Interpreting_chiral_spectra}. The question of which one of these methods is more effective in achieving em-chirality is well defined, and the measure of em-chirality allows to consistently compare different design options. 

The measure of the em-chirality of an object is based on the interaction of the object with fields of different helicity. These fields correspond to eigenstates of the helicity operator $\Lambda$ with eigenvalues $\pm 1$. Any eigenstate of $\Lambda$ with eigenvalue +1(-1) can be decomposed into a linear superposition of LCP(RCP) plane-waves. Moreover, one of the two Riemann-Silberstein-Birula combinations \cite{Birula1996} $\mathbf{E}\pm i Z\mathbf{H}$ is zero for an eigenstate of helicity \cite[Chap. 2,]{FerCorTHESIS}. In general, the computation of em-chirality requires the knowledge of the scattering matrix of the object. In other words, it requires the information about the interaction of the object with a complete set of fields of both helicities, like for example LCP and RCP plane-waves from a sufficiently large number of illumination directions \cite{Fruhnert2016b}. We now argue that, under normal illumination, and for a 2D array of identical elements with subwavelength periodicities, we only need four collinear plane waves for each frequency of interest, which, choosing $z$ as the axis and omitting the frequency, we label as
\begin{equation}
\left|\pm z,\lambda \right>,
\end{equation}
where the first label indicates the orientation of the momentum of the plane-wave along the $z$ direction, and the second label refers to the helicity eigenvalue of the plane-wave $\lambda = \pm 1$. 

We consider a 2D array of identical elements illuminated by a plane-wave whose momentum vector is perpendicular to the array plane. We assume that the array periodicities are smaller than the wavelength of the plane-wave. Under these conditions, the interaction can only produce plane-waves whose momentum is also perpendicular to the array plane, corresponding to the 0-th diffraction order. If the array lies in the $xy-$plane, then the four plane-waves $\left|\pm z,\pm \right>$ are a complete orthonormal basis for studying the light-matter interaction at each frequency. Any input/output field can be written as
\begin{equation}
\mathbf{E^{in}} = \sum{A^{in}_{\pm z,\pm} \left|\pm z,\pm \right>},\
\mathbf{E^{out}} = \sum{A^{out}_{\pm z,\pm} \left|\pm z,\pm \right>}.
\end{equation}

The interaction is fully characterized by the frequency dependent scattering matrix $\mathbf{S}$ of the structure, connecting the input and output fields. In the chosen basis we then write:
\begin{equation}
	\label{eq:S}
\begin{bmatrix} A^{out}_{+z,+}\\ A^{out}_{-z,+} \\ A^{out}_{+z,-} \\ A^{out}_{-z,-}\end{bmatrix}
= \mathbf{S}  \begin{bmatrix} A^{in}_{+z,+} \\ A^{in}_{-z,+} \\ A^{in}_{+z,-} \\ A^{in}_{-z,-}\end{bmatrix} 
=\begin{bmatrix} \mathbf{S_+^+} & \mathbf{S_-^+} \\ \mathbf{S_+^-} &  \mathbf{S_-^-}\end{bmatrix} \begin{bmatrix} A^{in}_{+z,+} \\ A^{in}_{-z,+} \\ A^{in}_{+z,-} \\ A^{in}_{-z,-}\end{bmatrix}. 
\end{equation}

In the last step of Eq.~(\ref{eq:S}) we have decomposed $\mathbf{S}$ into four different sub-matrices. Each sub-matrix links input fields of a specific helicity with output fields of a specific helicity. 

\begin{figure}[htbp]
\centering
\fbox{\includegraphics[width=0.98\linewidth]{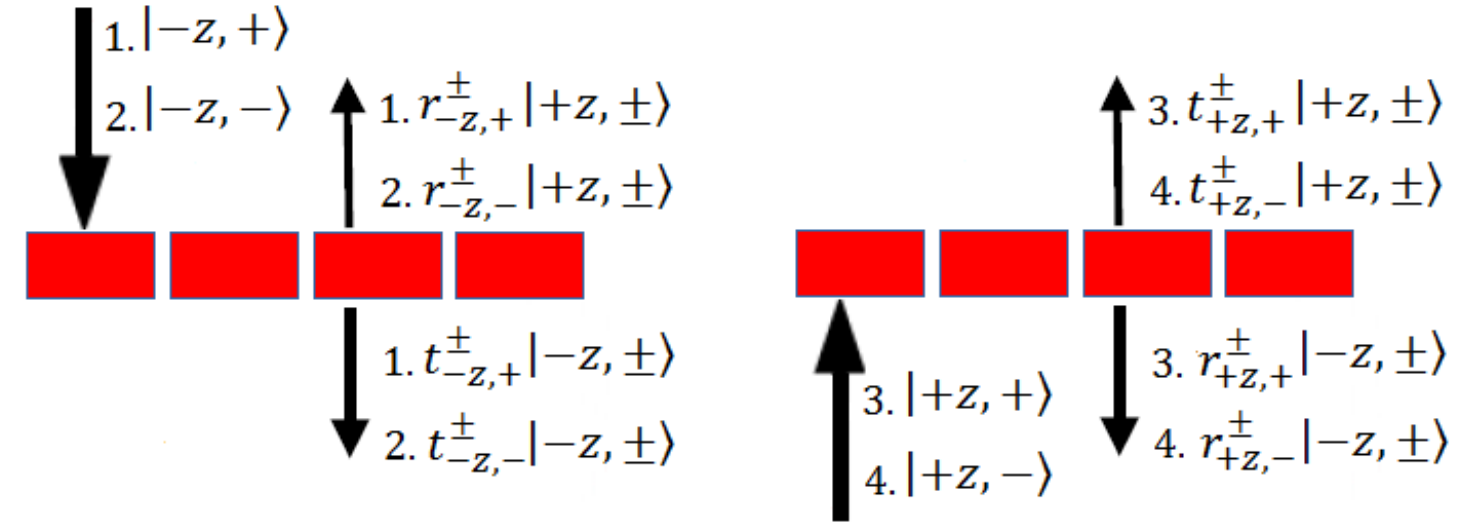}}
\caption{The structure will be illuminated with four different plane-waves coming from both $\pm$ $z$ directions and with both circular polarization handedness. The reflected and transmitted waves will then be decomposed into RCP plane-waves with complex amplitudes $r^-_{\pm z,\pm}$ and $t^-_{\pm z,\pm}$, and LCP plane-waves ($r^+_{\pm z,\pm}$,$t^+_{\pm z,\pm}$). Plane-waves are represented as $\left|\pm z,\pm \right>$, where the first label indicates the direction of propagation of the wave and the second label indicates its helicity.}
\label{fig:simulation_scheme}
\end{figure}

Figure~\ref{fig:simulation_scheme} summarizes the setting using the language of complex reflection and transmission coefficients $(r_{\pm z,\pm}^{\pm},\ t_{\pm z,\pm}^{\pm})$, where the $\pm$ subscript(superscript) refers to the helicity of the input(output) plane-wave, and the $\pm z$ subscript to the momentum of the input plane-wave. We can relate the complex reflection and transmission coefficients to elements of the $\mathbf{S}$ matrix,
\begin{equation}
	\label{eq:tands}
	\mathbf{S} = \begin{bmatrix}  
t_{+z,+}^+ & r_{-z,+}^+ & t_{+z,-}^+ & r_{-z,-}^+\\
r_{+z,+}^+ & t_{-z,+}^+ & r_{+z,-}^+ & t_{-z,-}^+ \\
t_{+z,+}^- & r_{-z,+}^- & t_{+z,-}^- & r_{-z,-}^- \\
r_{+z,+}^- & t_{-z,+}^- & r_{+z,-}^- & t_{-z,-}^- \end{bmatrix}.
\end{equation}

After the reflection and transmission coefficients are known, the em-chirality of the 2D array under normal illumination can be computed as follows. The definition of em-chirality \cite{Ivan:Maximum_chirality} uses the matrix $\mathbf{T} = \mathbf{S} - \mathbb{I}$, where $\mathbb{I}$ is the identity. We note that $\mathbf{T}$ is equal to the T-matrix of the object \cite{Mishchenko2016} times a factor of two that does not affect the value of the em-chirality.

Let us now denote by $\sigma\left(\mathbf{A}\right)$ the vector of singular values of the matrix $\mathbf{A}$, arranged from top to bottom in non-increasing order. We recall that the singular values are non-negative real numbers.

We use $\mathbf{T}$ to create two vectors: $\mathbf{v_+}$ and $\mathbf{v_-}$.
\begin{eqnarray}\label{eq:v+v-}
\mathbf{v_+} = \begin{bmatrix} \sigma\left(\mathbf{T^+_+}\right) \\ \sigma\left(\mathbf{T^-_+}\right) \end{bmatrix},\
\mathbf{v_-} = \begin{bmatrix} \sigma\left(\mathbf{T^-_-}\right) \\ \sigma\left(\mathbf{T^+_-}\right) \end{bmatrix}.
\end{eqnarray}
It is clear that $\mathbf{v_+}$($\mathbf{v_-}$) is obtained from the two sub-matrices which determine the interaction of the structure with LCP(RCP) illuminations. 

Finally, the measure of em-chirality $\chi$ is computed as the Euclidean distance between $\mathbf{v_+}$ and $\mathbf{v_-}$, normalized by the square root of the sum of the squared norms of $\mathbf{v_+}$ and $\mathbf{v_-}$
\begin{equation}
	\label{eq:x}
\chi = \frac{|\mathbf{v_+}-\mathbf{v_-}|}{\sqrt{|\mathbf{v_+}|^2+|\mathbf{v_-}|^2}} = \frac{\sqrt{\sum_i{\left(v_{+,i}-v_{-,i}\right)^2}}}{\sqrt{\sum_i{v_{+,i}^2}+\sum_i{v_{-,i}^2}}}.
\end{equation}

Arguments identical to those in \cite{Ivan:Maximum_chirality} show that this measure is upper bounded by 1: $\chi\in[0,1]$. We note that the procedure is the same for any single direction of incidence.

Before computing the em-chirality of different arrays, we analyze how different symmetries impose restrictions on the elements of $\mathbf{T}$, and how these restrictions affect the em-chirality of the arrays.  

A 2D array in the $xy-$plane can only be symmetric for mirror reflections across planes that either contain the $z$ axis or are perpendicular to it. The possible angles of discrete rotational symmetry along the $z$ axis are also restricted in periodic 2D arrays. For example, they must be multiples of 90 degrees for a square lattice and of 60 degrees for a hexagonal lattice. Additionally, reciprocity is always met in light-matter interactions unless the time-reversal symmetry is broken by, for example, an external biasing magnetic field or a time dependent perturbation.

Table \ref{tab:geometry_conditions} contains the restrictions imposed on the $\mathbf{S}$ matrix by some of the aforementioned symmetries. The same restrictions apply to the $\mathbf{T}$ matrix. 

\begin{table}[htbp]
\centering
\caption{Restrictions imposed on the scattering matrix by reciprocity and geometrical symmetries. }
\begin{tabular}{ccc}
\hline
Symmetry & Restriction \\
\hline
$Reciprocity$ & $\left<\eta _2,\lambda _2\right| \mathbf{S} \left| \eta _1,\lambda _1\right> = \left<-\eta _1,\lambda _1\right| \mathbf{S} \left| -\eta _2,\lambda _2  \right>$ \\
$M_z$ & $\left<\eta _2,\lambda _2\right| \mathbf{S} \left| \eta _1,\lambda _1\right> = \left<-\eta _2,-\lambda _2\right| \mathbf{S} \left| -\eta _1,-\lambda _1  \right>$ \\
$M_{x(y)}$ & $\left<\eta _2,\lambda _2\right| \mathbf{S} \left| \eta _1,\lambda _1\right> = \left<\eta _2,-\lambda _2\right| \mathbf{S} \left| \eta _1,-\lambda _1  \right>$ \\
$R_z(\frac{2\pi}{n})$ & $\left<\eta _2,\lambda _2\right| \mathbf{S} \left| \eta _1,\lambda _1\right> = e^{i\frac{2\pi}{n}\left(\lambda _2-\lambda _1\right)}\left<\eta _2,\lambda _2\right| \mathbf{S} \left| \eta _1,\lambda _1\right>$ \\
$R_{x(y)}(\pi)$ & $\left<\eta _2,\lambda _2\right| \mathbf{S} \left| \eta _1,\lambda _1\right> = \left<-\eta _2,\lambda _2\right| \mathbf{S} \left| -\eta _1,\lambda _1  \right> $\\
\hline
\end{tabular}
The notation $\left<\eta _2,\lambda _2\right| \mathbf{S} \left| \eta _1,\lambda _1\right>$ denotes the matrix element connecting the input $| \eta _1,\lambda _1\rangle$ to the output $| \eta _2,\lambda _2\rangle$, where $\eta_i$ refers to the direction of the plane-waves and $\lambda_i$ to their helicities. $M_\alpha$ is the mirror reflection across the plane perpendicular to axis $\alpha$, and $R_\alpha(\beta)$ a rotation along axis $\alpha$ by an angle $\beta$.
  \label{tab:geometry_conditions}
\end{table}

\begin{figure}[htbp]
\centering
\fbox{\includegraphics[width=0.80\linewidth]{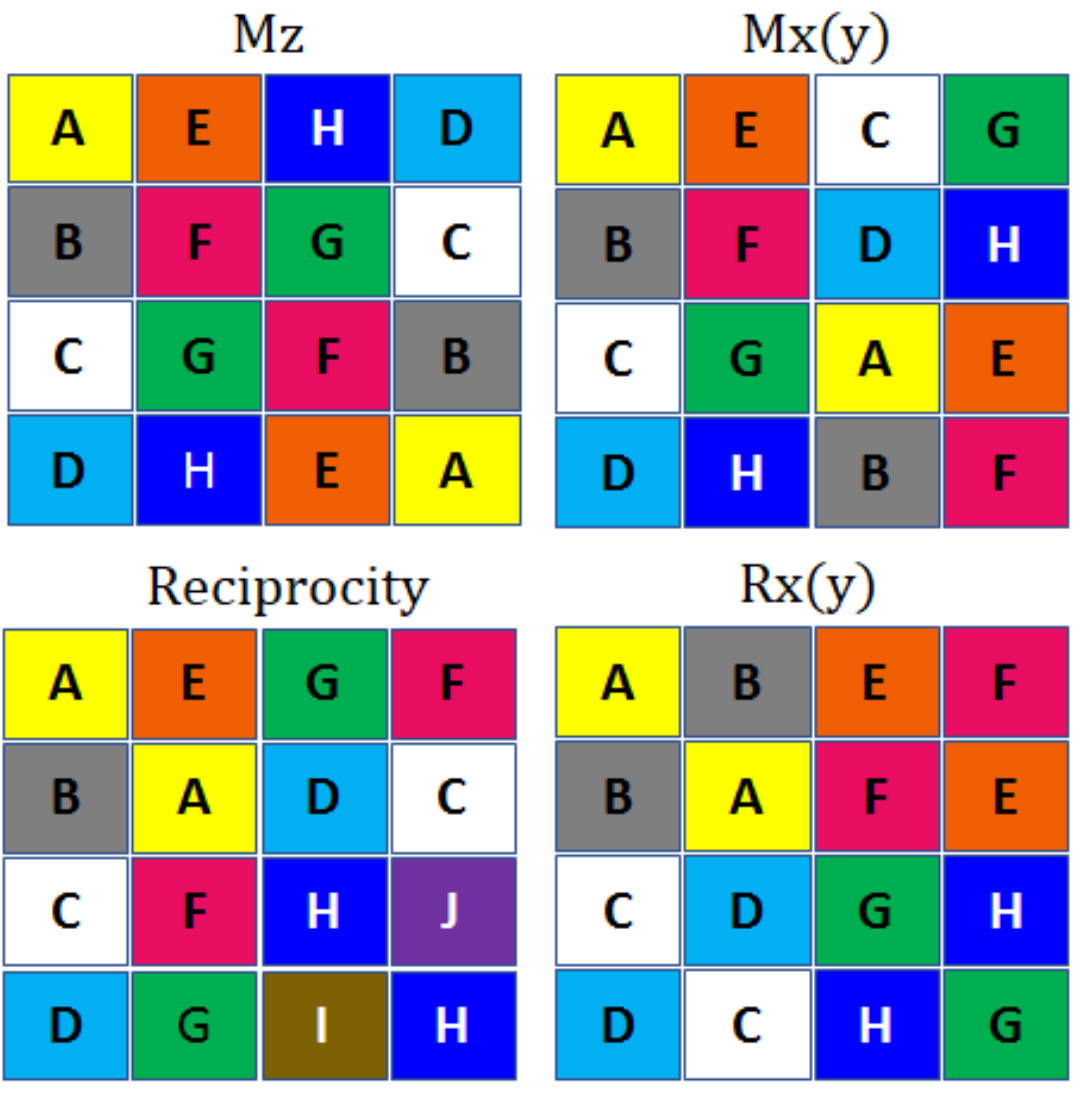}}
\caption{Symmetry restrictions on the $\mathbf{S}$ matrices of 2D arrays.}
\label{fig:M_symmetries}
\end{figure}

In Fig.~\ref{fig:M_symmetries} we show the pattern of the scattering matrix imposed by mirror symmetries, rotational symmetries, and reciprocity. Using Fig.~\ref{fig:M_symmetries} and Eqs.~\ref{eq:v+v-} and \ref{eq:x}, one can easily deduce that any 2D array presenting a mirror symmetry will be electromagnetically achiral, confirming the consistency of the measure with the geometrical definition of chirality. For example, the pattern imposed by $M_{x(y)}$ is that $\mathbf{T^+_+}=\mathbf{T^-_-}$ and $\mathbf{T^-_+}=\mathbf{T^+_-}$, then $\mathbf{v_+}=\mathbf{v_-}$ and $\chi=0$. From the pattern imposed by $M_z$, one can see that the matrices for input helicity -1 are a unitary transformation of the matrices for input helicity +1
\begin{equation}
	\mathbf{T^-_-}= \begin{bmatrix} 0&1\\1&0 \end{bmatrix}\mathbf{T^+_+}\begin{bmatrix} 0&1\\1&0 \end{bmatrix},\ \mathbf{T^+_-}= \begin{bmatrix} 0&1\\1&0 \end{bmatrix}\mathbf{T^-_+}\begin{bmatrix} 0&1\\1&0 \end{bmatrix},
\end{equation}
which also implies $\mathbf{v_+}=\mathbf{v_-}$, because the singular values of a matrix are invariant under unitary transformations. Similarly, one can see that no rotational symmetry by itself forces a structure to be electromagnetically achiral, and that only the helicity preserving interaction ($\mathbf{T_+^+}$ and $\mathbf{T_-^-}$) contribute to the degree of chirality for any reciprocal array.

\begin{figure}[h]
\centering
\fbox{\includegraphics[width=0.9\linewidth]{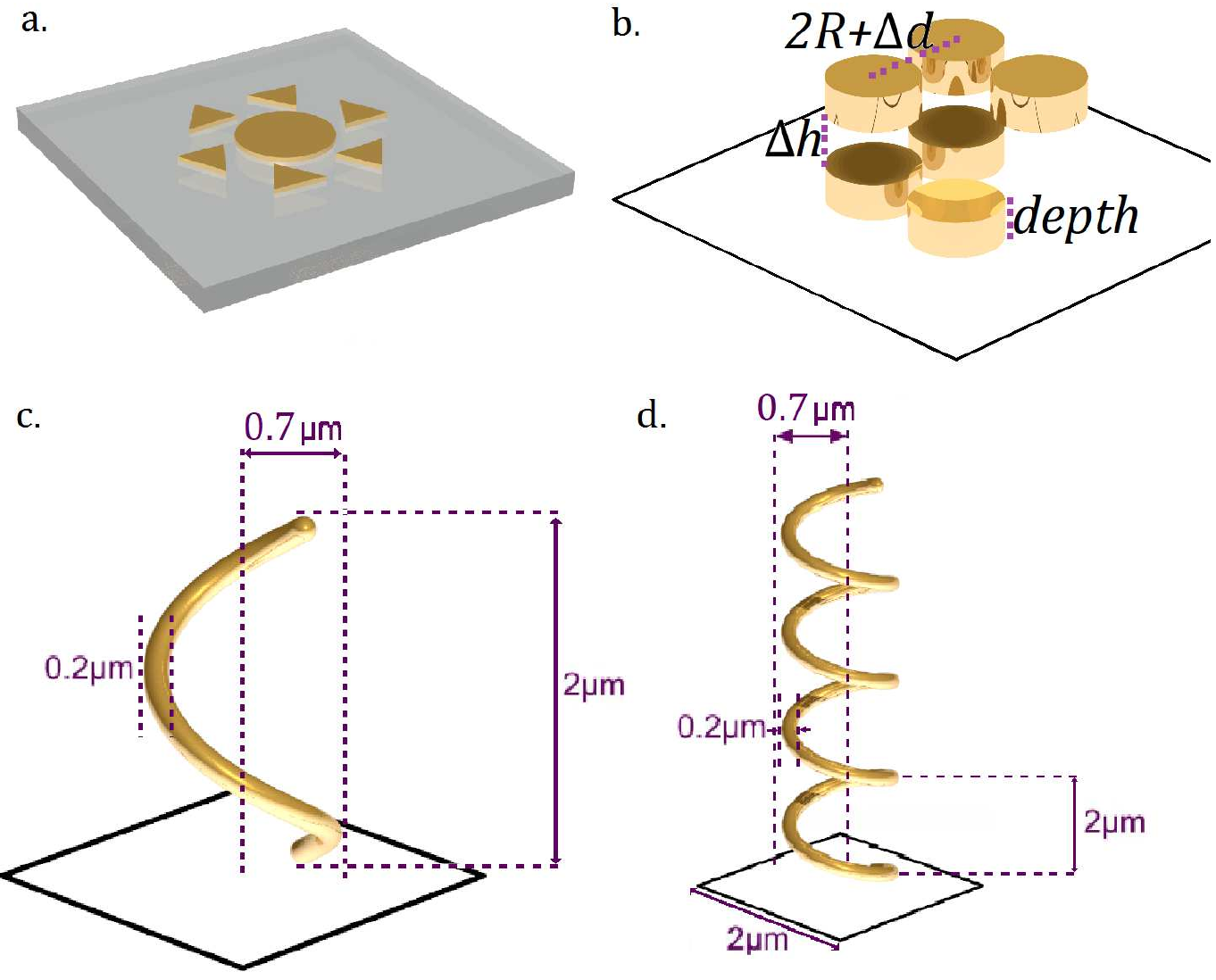}} %
\caption{Studied structures. Structure (a) is composed of a disc with a radius of 490 nm and six triangles placed around the disc. The triangles have a base of 420 nm and a height of 700 nm. The distance between the center of the triangles and the center of the disc is 1190 nm. All the elements have a depth of 140  nm. The structure is placed on top of a 2 $\mu$m glass substrate (n = 1.5). Structure (b) is composed of six gold discs arranged as shown in the picture. The parameter values are: R = 350 nm, $\Delta$d = 140 nm, $\Delta$h = 420 nm, depth = 280 nm. Structure (c) and structure (d) are two helices with 1 and 4 turns respectively. The parameters of the helix are the ones specified in \cite{Gansel:Gold_helix}. All the structures are made of gold and present a square lattice with a periodic length of L = 3500 nm. Gold refractive index is calculated using the Drude model with the parameters specified in \cite{Gansel:Gold_helix}.}
\label{fig:structures_example}
\end{figure}

Finally, we compute the frequency dependent em-chirality of four different 2D periodic squared arrays for wavelengths ranging from to 3.51 to 8 $\mu$m (85.4 to 37.5 THz). Figure \ref{fig:structures_example} shows the unit cells of the arrays and their periods. 

Structure (a) is a scaled up version of the one reported in \cite{Hopkins:Circular_dichroism}. It consist of a central disc surrounded by six isosceles triangles. According to the previous discussion, the corresponding 2D array would be em-achiral because of its $M_z$ mirror symmetry. We break this symmetry by placing the structure on top of a glass substrate. Structure (b) is a scaled up version of the one studied in \cite{Schaferling:Enhanced_optical_chirality}. It is composed of two L-shaped arrangements of three gold nano-discs that are placed on top of each other with a 90 degree twist. The last two structures, studied in \cite{Gansel:Gold_helix}, are two gold helices with different number of turns (1 and 4), and otherwise identical parameters. 
\begin{figure}[h]
\centering
\includegraphics[width=\linewidth]{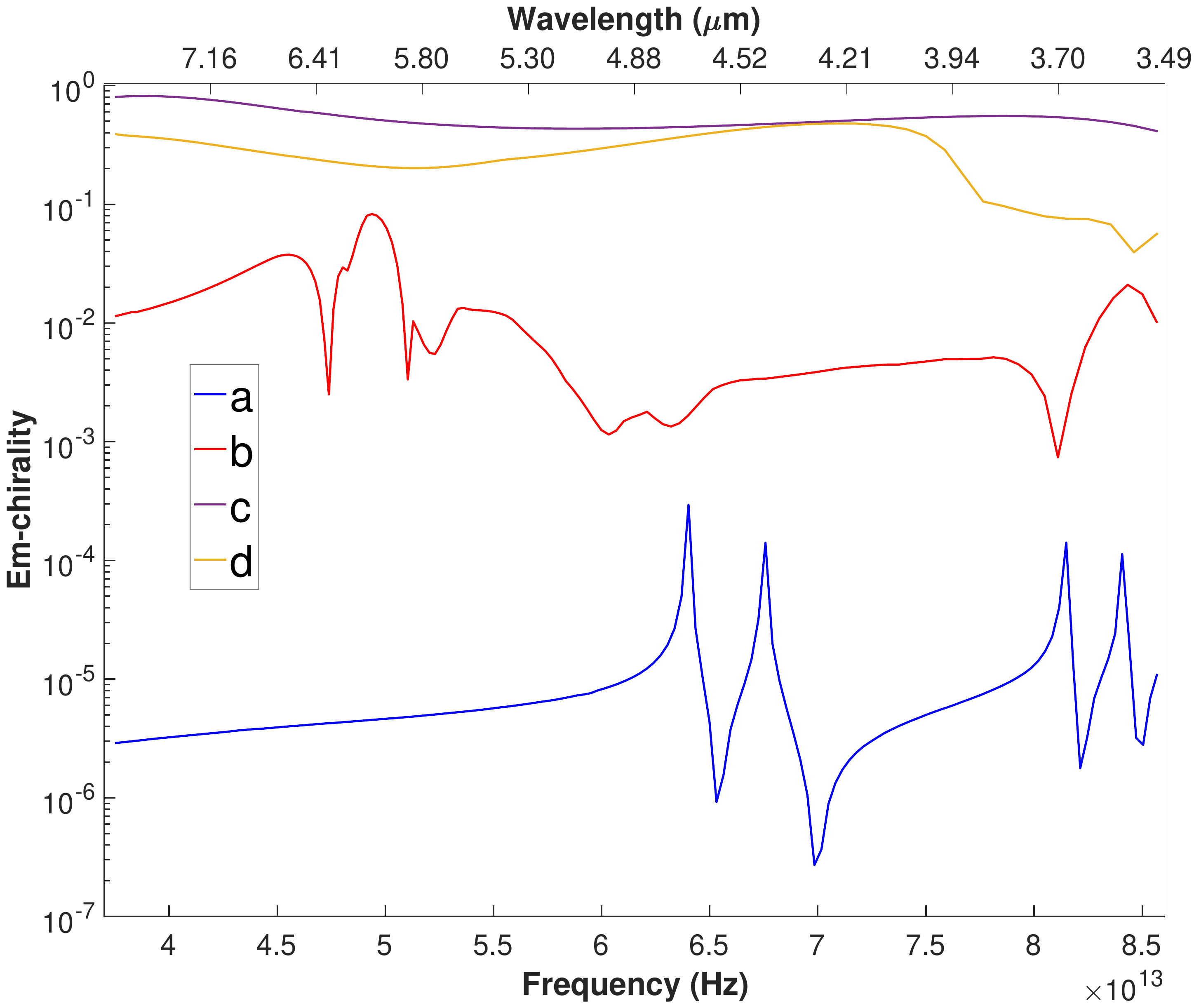}
\caption{Em-chirality of the four arrays described in Fig.~\ref{fig:structures_example}. The maximum values for each case (a to d) are: 0.0029, 0.08, 0.82, and 0.48.}
\label{fig:numerical_results}
\end{figure}

The frequency dependent em-chirality of the arrays is computed as follows. First, their reflection and transmission coefficients are obtained using the FEM solver JCMsuite \cite{JCMsuite}. The em-chirality is then computed following the procedure outlined above, with Eq.~(\ref{eq:tands}) as the starting point. The results are shown in Fig.~\ref{fig:numerical_results}. Notably, the ordering established by the em-chirality of these arrays is frequency independent.

The least em-chiral array is the one with structure (a) as its unit cell. We recall that its em-chirality can be non-zero only because of the presence of the glass substrate. In this respect, the location of the peaks(dips) of its em-chirality can be interpreted as the frequencies where the relevance of the substrate on the response of the array is locally maximal(minimal). We have obtained very similar em-chirality magnitudes for substrates thicknesses of 200 nm and 5 $\mu$m. The low em-chirality values are consistent with the results in \cite{Schaferling:Enhanced_optical_chirality}, where it was observed that structures with variation in the third dimension seem to be necessary in order to get high values of the near-field's chirality density. The authors of \cite{Schaferling:Enhanced_optical_chirality} then proposed the structure in Fig.~\ref{fig:structures_example} (b). For our (scaled up) version, the em-chirality of the corresponding array is between one and three orders of magnitude larger than for structure (a). The highest values are, however, obtained for the two helices, and the single turn helix is more em-chiral than the four turn one for all frequencies. The extremal interaction between circularly polarized radiation and metal helices has been studied for decades now \cite{Wheeler1947,Semchenko2009,Gansel:Gold_helix,Karilainen2012b}. Our results indicate that helices are a reference/canonical structure for achieving large em-chirality. 

In conclusion, we have given a detailed recipe for computing the em-chirality for 2D arrays of subwavelength periodicity under normal illumination. This allows the unambiguous ordering and comparison of such arrays according to their em-chirality. We have analyzed the consequences that different possible symmetries of the array have on its em-chirality. In particular, we have shown that if the array has any mirror symmetry its em-chirality is zero. The frequency dependent em-chirality of four different arrays shows that, among the structures studied here, helices are the most suitable objects for achieving high em-chirality, and indicates that the presence of substrates is not an efficient mechanism for breaking the transverse mirror symmetry of planar 2D arrays. The measure will assist in the design of highly em-chiral arrays which, due to their extreme light matter interaction properties, have applications in e.g. polarization control in both far- and near-fields, and sensing of chiral molecules.

\textbf{Funding.} Deutsche Forschungsgemeinschaft (DFG) grant CRC 1173; Marie Curie ITN EID project "NOLOSS" grant 75745; Karlsruhe School of Optics and Photonics (KSOP).

\bibliography{References_paper}

\clearpage
\bibliographyfullrefs{References_paper}

\end{document}